\documentstyle[12pt]{article}
\begin{document}
\thispagestyle{empty}
\vspace*{0.2in}
\begin{flushright}
\hfill\vbox{\hbox{UA/NPPS-1-03}}
\end{flushright}
\vspace{2cm}
%\begin{center}
%\documentstyle[12pt]{article}
%\textheight 9in \textwidth 6.5in \oddsidemargin -.2in \topmargin -.5in
%\renewcommand {\baselinestretch}{1.5}
%\begin{document}
\begin{center}
{\large{\bf A Space-time approach to multi-gluon loop computations in QCD:\\ 
An application to effective action terms}}\\
%{\bf WORLD-LINE TECHNIQUES FOR RESUMMING\\
%GLUON RADIATIVE CORRECTIONS}\\
\vspace{1.5cm}
A. I. Karanikas
\footnote{akaranik@cc.uoa.gr}~~and~C. N. Ktorides\footnote{cktorid@cc.uoa.gr}\\
\smallskip
\textit{University of Athens, Physics Department\\
Nuclear \& Particle Physics Section\\
Panepistimiopolis, Ilisia GR 157--71, Athens, Greece}
\end{center}
\vspace{3.5cm}
%\numberwithin{equation}{section}
%%%%%%%%%%%%%%%%%%%%%%%%%%%%%%%%%%%%%%%%%%%%%%%%%%%%%%%
\begin{abstract}
The applicability of the space-time formulation of the gluonic sector of QCD in terms
of the Polyakov worldline path integral, via the use of the background field
gauge fixing method, is extended to multi-gluon loop configurations. Relevant master
formulas are derived for the computation of effective action terms. 
\end{abstract} 
\newpage

Space-time based approaches to QCD, employing first quantization methods, have 
established a viable alternative for expediting perturbative computations in the 
theory. The first attempts in this direction have 
used strings as underlying agents [1-3] through which a set of 
prescriptions, known as Bern-Kosower rules, were successfully applied at 
the one gluon loop level. Subsequently, Strassler [4] established the 
possibility to reproduce the whole approach by basing the formalism 
on worldline paths instead of strings, see also Refs [5,6]. More recently, 
Sato, Schmidt and collaborators [7,8] have extended the worldline considerations 
to the two loop level. In this letter we apply  
Polyakov's version [9] of the worldline casting 
of QCD, which we have been consistently pursuing for some time [10-12], 
to multi-gluon loop configurations and 
formulate specific rules for constructing the relevant expressions 
that represent them. More than just as a strategy pertaining to perturbative 
calculations this approach provides a framework within which a systematic treatment 
of the propagation of particle modes accommodating Wilson lines (loops) can be achieved.

As in a recent paper [12], in which we presented one gluon loop master formulas for effective 
action terms, we shall formulate the worldline path integral for the gluon by 
employing the background gauge fixing 
method [13], which is based on a splitting of the gauge field 
potentials $A_\mu^a$ by setting $A_\mu^a=a_\mu^a+B_\mu^a$ and treating 
the  $a_\mu^a$ modes as fully fluctuating in a background 
represented by the fields $B_\mu^a$. For the present purposes the latter will 
be considered as classical. The idea, however, that in a quantized context they could 
be used as `carriers' of non-perturbative 
physics in QCD presents interesting possibilities on which we shall devote 
some comments in the end. In any case, they facilitate the gauge fixing through the condition

\begin{equation}
(\delta^{ab}\partial_\mu +gf^{abc}B^c_\mu)\alpha^b_\mu\equiv D_\mu^{ab}(B)\alpha_\mu^b=0.
\end{equation}

Our starting point is the generating functional (partition function in Euclidean
space-time) given, for the gluonic sector of the theory, by
\begin{eqnarray}
Z(B)&=&N\int{\cal D}\bar{c}{\cal D}c{\cal D}\alpha exp(-S_A)=NZ_0Det^{-{1\over 2}}
[-(D^2)^{ab}\delta_{\mu\nu}-2gf^{abc}F_{\mu\nu}^c(B)]
Det[-R({\delta\over\delta J})]\nonumber\\&&\times exp\left[-S_I\left({\delta\over\delta J}\right)\right] 
exp\left[{1\over 2}\int d^4x\,d^4x' J_\mu^a(x)iG^{[1]}(x,x')^{ab}_{\mu\nu} J^b_\nu
\right]_{J=0},
\end{eqnarray}
where $c,\bar{c}$ denote ghost fields,
$S_I$ is the interaction part of the Yang-Mills action for the $\alpha$ fields,
$Z_0$ is the classical contribution, $iG^{[1]}$ is the gluon propagator and
\begin{equation}
R^{ab}({\alpha})=(D^2)^{ab}+g\overleftarrow{D}_\mu^{ac}f^{cbd}\alpha^d_\mu.
\end{equation}

Propagators of particle modes (field quanta) according to the Polyakov worldline path integral 
have the following generic form, see Refs [10-12], 
\begin{equation}
iG^{[s]}(x,x')_{mn}^{ab}=\int\limits_0^\infty\,dT\int\limits_{x(0)=x',x(T)=x}Dx(t) 
exp\left({-1\over 4}
\int\limits_0^Tdt\,\dot{x}^2\right)\Phi^{[s]}_{mn}[C^{xx'}]U[C^{xx'}]^{ab},
\end{equation}
where $s=0,1/2,1,...$ stands for the spin of the propagating particle entity and $m,n$
are `Lorentz generator' indices for each given value of $s$, e.g.,
$mn \rightarrow \mu\nu$ for spin 1. Furthermore, $\Phi^{[s]}_{mn}[C^{xx'}]$ is the 
so-called spin-factor for the particle mode whose generic form is ($P$ stands for path ordering)
\begin{equation}
\Phi^{[s]}_{mn}[C^{xx'}]=Pexp\left({i\over 2}\int\limits_0^Tdt\,J\cdot\omega\right)_{mn}
;\quad x(0)=x,\,x(T)=x'
\end{equation}
with $\omega_{\rho\sigma}$ the tensor ${T\over2}[\ddot{x}_\rho(t)\dot{x}_\sigma(t)-
\ddot{x}_{\sigma}(t)
\dot{x}_\rho(t)]$ whose indices are contracted with those of the generators, namely
$J^{\rho\sigma}_{mn}$ and where $U[C^{xx'}]^{ab}$ is the Wilson line element, specified 
in terms of the background gauge fields, i.e.
\begin{equation}
U[C^{xx'}]^{ab}=Pexp\left(ig\int\limits_0^Tdt\,\dot{x}(t)\cdot B\right)^{ab}.
\end{equation}

In the particular case of the pure gluonic sector, to which we shall exclusively devote 
our attention 
in this paper, we have $(J_{\mu\nu})^{\rho\sigma}\rightarrow(\delta_{\rho\mu}\delta_{\sigma\nu}
-\delta_{\rho\nu}\delta_{\sigma\mu})$, while for the propagation of the ghost modes the 
spin factor is unity. For completeness, we mention that in the case of spin-1/2 modes 
one has $(J_{mn})^{\rho\sigma}\rightarrow {1\over 2}
\sigma_{\alpha\beta}^{\rho\sigma}$,
where the $\alpha,\,\beta$ are Dirac spinor indices.

The above set of formulas suffice to conduct any perturbative, at least, computation in
the theory. To be concrete, we can produce prescriptions  for combining the worldline
carriers of gluons and ghosts thereby arriving at the worldline version of the Feynman 
rules for the corresponding vertices.
Given that the gist of any such computation are the 1PI configurations, we
shall, in the present paper, restrict our considerations to effective action functionals,
whose one-loop contribution term has been confronted in Ref  [12].
For two-loops we must separately deal with the configuration having
a single 4-gluon vertex and the one with two 3-gluon vertices. For the first case we
obtain, upon inserting the Polyakov worldline path integral expression for the   
propagators involved and 
designating by $C_1$ and $C_2$ the respective loops that compose the `figure-8' formation,
\begin{eqnarray}
\Gamma_4^{(2)}&=&{1\over 8}\int d^4x\left[\prod\limits_{i=1}^2
\int\limits_0^\infty\,dT_i\right]
\left[\prod\limits_{i=1}^2\int\limits_{x_i(0)=x_i(T_i)=x}Dx_i(t_i)exp\left(-{1\over 4}
\int\limits_0^{T_i}\,dt_i\,\dot{x}_i^2\right)\right]\nonumber\\&&\times
\Phi^{[1]}_{\mu\nu}[C_2^{xx}]\Phi^{[1]}_{\rho\sigma}[C_1^{xx}]
(V_4)^{abcd}_{\mu\nu\rho\sigma}U[C_2^{xx}]^{ab}U[C_1^{xx}]^{cd},
\end{eqnarray}
with
\begin{eqnarray}
(V_4)^{abcd}_{\mu\nu\rho\sigma}&=&g^2[f^{eab}f^{ecd}(\delta_{\mu\rho}\delta_{\nu\sigma}
-\delta_{\mu\sigma}\delta_{\nu\rho})+
f^{eac}f^{ebd}(\delta_{\mu\nu}\delta_{\rho\sigma}
-\delta_{\mu\sigma}\delta_{\nu\rho})\nonumber\\&&\quad\quad
+f^{ead}f^{ebc}(\delta_{\mu\nu}\delta_{\rho\sigma})-\delta_{\mu\rho}\delta_{\nu\sigma})]
\end{eqnarray}
the well known 4-gluon vertex.

In order to deal with the second configuration, which involves a derivative 
operation at each vertex, we first observe that
\begin{equation}
D^{ab}_{\mu,x}iG^{[s]}(x,x')^{bc}_{\nu\rho}={1\over 2}\int\limits_0^\infty dT
\int\limits_{x(0)=x',x(T)=x}Dx(t) \dot{x}_\mu(T)
exp\left(-{1\over 4}
\int\limits_0^Tdt\,\dot{x}^2\right)\Phi^{[1]}_{\nu\rho}[C^{xx'}]U[C^{xx'}]^{ac}
\end{equation}
and similarly, with a relative minus sign, when the covariant derivative acts with
respect to $x'$. 
We now face a situation where the relevant configuration has three,
open line, branches. The following organization is adopted. Place $x$ and$x'$
on diametrically opposite points on a circle. Go, always from $x'$ to $x$, in three different
ways. Once along one semicrcle, contour $C_1$, once along the diameter, contour $C_2$,
and once along the other semicircle, contour $C_3$. Following the convention that velocities
are positive at the end point we determine, after some algebraic manipulations,
\begin{eqnarray}
&&\Gamma_3^{(2)}={1\over 2}{1\over 3!}\int d^4xd^4x'\left[\prod\limits_{i=1}^3
\int\limits_0^\infty\,dT_i\right]
\left[\prod\limits_{i=1}^3\int\limits_{x_i(0)=x',x_i(T_i)=x}Dx_i(t_i)exp\left(-{1\over 4}
\int\limits_0^{T_i}\,dt_i\,\dot{x}_i^2\right)\right]\nonumber\\&&\quad\quad\times
\Phi^{[1]}_{\mu\rho}[C_3^{xx'}]\Phi^{[1]}_{\nu\sigma}[C_2^{xx'}]\Phi^{[1]}_{\kappa\lambda}
[C_1^{xx'}](V_3^x)^{abc}_{\mu\nu\kappa}(V_3^{x'})^{def}_{\rho\sigma\lambda}
U[C_3^{xx'}]^{ad}U[C_2^{xx'}]^{be}U[C_1^{xx'}]^{cf},
\end{eqnarray}
where the first vertex reads
\begin{eqnarray}
(V_3^x)^{abc}_{\mu\nu\kappa}&=&{i\over 2}gf^{abc}\{\delta_{\mu\nu}[\dot{x}_2(T_2)
-\dot{x}_1(T_1)]_\kappa+\delta_{\mu\kappa}[\dot{x}_1(T_1)-\dot{x}_3(T_3)]_\nu \nonumber\\&&
\quad\quad+\delta_{\nu\kappa}[\dot{x}_3(T_3)-\dot{x}_2(T_2)]_\mu\},
\end{eqnarray}
while the second has a relative minus sign -accounting for opposite velocity orientations- 
and makes the replacement $T_i\rightarrow 0$. 

The last specification needed is for the the two loop configuration 
which involves vertices with ghosts. Choosing our paths the same as for 
$\Gamma_3^{(2)}$, with $C_1$ and $C_2$ traversed by ghosts and $C_3$ by gluons and once 
determining the action of the corresponding covariant derivative operators on the ghost field
propagator, in analogy to Eq. (9), the following result is obtained
\begin{eqnarray}
\Gamma_{gh}^{(2)}&=&{1\over 2}{1\over 3!}\int d^4xd^4x'\left[\prod\limits_{i=1}^3
\int\limits_0^\infty\,dT_i\right]
\left[\prod\limits_{i=1}^3\int\limits_{x_i(0)=x',x_i(T_i)=x}Dx_i(t_i)exp\left(-{1\over 4}
\int\limits_0^{T_i}\,dt_i\,\dot{x}_i^2\right)\right]\nonumber\\&&\times
\Phi^{[1]}_{\mu\nu}[C_3^{xx'}]\Phi^{[0]}[C_2^{xx'}]\Phi^{[0]}
[C_1^{xx'}](V_g^x)^{abc}_{\mu}(V_g^{x'})^{def}_\nu
U[C_3^{xx'}]^{ad}U[C_2^{xx'}]^{be}U[C_1^{xx'}]^{cf},
\end{eqnarray}
where
\begin{equation}
(V_g^x)^{abc}_{\mu}=-{i\over 2}f^{abc}\dot{x}_{1\mu}(T_1)
\end{equation}
with $1\rightarrow 2$ and $T_1\rightarrow 0$ for the second vertex.

One observes that the expressions for the vertices are related to those of the second
quantized (field theoretical) formulation of QCD via the correspondence
$p\leftrightarrow {i\over 2}\dot{x}$. This occurence explicitly illustrates the
space-time character of our approach to the theory.

From what we have so far  presented the following rules follow for the space-time approach
to perturbation theory:

$\bullet${Draw all diagrams (1PI for our present purposes) in terms of 
worldlines for the particle modes which employ the allowed vertices. The analytical 
expressions for the latter are furnished by Eqs. (8), (11) and (13).}

$\bullet$ {For each line segment between adjacent vertices assign the appropriate 
spin factor and Wilson line(loop).}

$\bullet$ {Each diagram is assigned a combinatorial factor which coincides with that
of the corresponding conventional Feynman diagram.}

The perturbative computation of the effective action, at two loop level, proceeds once we 
expand the Wilson exponential factors entering the corresponding expressions. Keeping
the Mth order, in $g$, of the expansion means that one determines up to M+2-order 
contributions to the effective action. Before taking this step we find it convenient to 
recast our two loop expressions in a form wherein the various worldlines entering them
are described in terms of parameters running along the traversed (Wilson) paths  as
opposed to being characterized through the points of reference $x,x'$. 
In this way one circumvents zero mode problems due to translational invariance while
at the same time achieves a form which, in the Feynman diagrammatic language, amounts
to directly reaching expressions at the stage where Feynman parameters have been introduced.

The new form of $\Gamma_4^{(2)}$ becomes, once making the redefinition
$t_2\rightarrow T_1+T_2-t_2$ and the replacements $T_1=s_1$, $T_1+T_2=s$, 
\begin{eqnarray}
\Gamma_4^{(2)}&=&{1\over 8}\int\limits_0^\infty ds \int\limits_0^s ds_1
\int Dx(t)\delta[x(s)-x(0)]\delta[x(s)-x(s_1)]exp\left(-{1\over 4}
\int\limits_0^s dt\,\dot{x}^2\right)\nonumber\\&&\quad\quad\times
\Phi^{[1]}_{\mu\nu}[C^{ss_1}]\Phi^{[1]}_{\rho\sigma}[C^{s_10}]
(V_4)^{abcd}_{\mu\nu\rho\sigma}U[C^{ss_1}]^{ab}U[C^{s_10}]^{cd},
\end{eqnarray}
where we have used the notation $\Phi^{[1]}_{\mu\nu}[C^{ss'}]\equiv 
Pexp\left({i\over 2}\int\limits_s^{s'}dt\,J\cdot\omega\right)_{\mu\nu}$ and similarly for
$U[C^{ss'}]^{ab}$. Also, we have taken into account the fact that the contour segment $C_2$ of
Eq (7) is now being reversely traversed since, on account of the uniform parametrization,
one branch is given a clockwise and the other a counterclockwise orientation.

For the configurations with the pair of 3-vertices the called for changes are
$t_3\rightarrow T_1+T_2+t_3$ and $t_2\rightarrow T_1+T_2-t_2$ while the corresponding
replacements are $T_1=s_1$, $T_1+T_2=s_2$ and $T_1+T_2+T_3=s$. Once taking into account
direction reversals since, for the sake of achieving
a single parametrization, the contour is to be traced in a continuous manner, we obtain   
\begin{eqnarray}
\Gamma_3^{(2)}&=&{1\over 8}\int\limits_0^\infty ds \int\limits_0^s ds_2
\int\limits_0^{s_2} ds_1
\int Dx(t)\delta[x(s)-x(s_1)]\delta[x(0)-x(s_2)]exp\left(-{1\over 4}
\int\limits_0^s\,dt\,\dot{x}^2\right)\nonumber\\&&\quad\quad\times
V_3[C,\dot{x}]f^{abc}f^{def}U[C^{ss_2}]^{ad}U[C^{s_1s_2}]^{be}U[C^{s_10}]^{cf},
\end{eqnarray}
where 
\begin{eqnarray}
&&\quad\quad\quad V_3[C,\dot{x}]\equiv g^2(\{\Phi^{[1]}_{\mu\nu}[C^{ss_2}]
Tr_L\Phi^{[1]}[C^{s_20}]-\Phi^{[1]}_{\mu\rho}[C^{ss_2}]\Phi^{[1]}_{\nu\rho}[C^{s_20}]\}
\dot{x}_\mu(s_1-0)\nonumber\\&&\times[\dot{x}_\nu(0)+\dot{x}_\nu(s_2-0)]
+\{\Phi^{[1]}_{\rho\mu}[C^{ss_1}]\Phi^{[1]}_{\rho\nu}[C^{s_10}]+\Phi^{[1]}_{\mu\nu}[C^{s0}]\}
\dot{x}_\mu(s_1-0)\dot{x}_\nu(0))
\end{eqnarray}
and similarly for $\Gamma_{gh}^{(2)}$ with
\begin{equation}
V_3[C,\dot{x}]\rightarrow
V_g[C,\dot{x}]\equiv -g^2\Phi^{[1]}_{\mu\nu}[C^{ss_2}]\dot{x}_\mu(s_1-0)\dot{x}_\nu(s_1+0)).
\end{equation}

We now proceed with the expansion of the Wilson exponential. To Mth order we have
\begin{eqnarray}
U[C^{ss'}]^{ab}&=&\delta^{ab}+ig(t_G^{c_1})^{ab}\int\limits_s^{s'} dt_1\dot{x}(t_1)\cdot B^{c_1}
[x(t_1)]+\cdot\cdot\cdot \nonumber\\&&
+\left[\prod\limits_{n=M}^1ig(t^{c_n}_G)\right]^{ab}
\prod\limits_{n=M}^1\int\limits_s^{s'}\theta(t_n-t_{n-1})\dot{x}(t_n)\cdot B^{c_n}[x(t_n)]
+{\cal O}(g^{M+1}).
\end{eqnarray}
Upon writing
\begin{equation}
B^a_\mu=\int\frac{d^Dq}{(2\pi)^D}e^{iq\cdot x}\tilde{\epsilon}_\mu^a(q)
\end{equation}
and using the equation of motion $D^{ab}_\mu F^b_{\mu\nu}(B)=0$ obeyed by the
(classical) background field, one determines
\begin{equation}
\tilde{\epsilon}^a_\mu(q)=(2\pi)^D\delta(q-p_a)\epsilon^a_\mu+g(2\pi)^D\delta(q-p_b-p_c)
(t^a_G)^{bc}2\frac{\epsilon^b\cdot p_c\epsilon^c_\mu}{(p_b+p_c)^2}+{\cal O}(g^2),
\end{equation}
where the $\epsilon^a_\mu$ are polarization vectors and the $p_{a\mu}$ four
momenta for the background fields. In a generic computation of an n-point Green's function
the above expansion serves as the means by which a given worldline gluon 
(multi)loop configuration is `punctured' by a network of gluonic tree diagrams whose 
contribution to the overall computation is determined by classical field perturbation theory. 
For the purpose of computing contributions to the effective action we are exclusively
interested in 1PI terms, hence it suffices to set $\{B^{a_n}_{\mu_n}[x(t_n)]\}\rightarrow
\{\epsilon^n_{\mu_n}exp[ip_n\cdot x(t_n)]\}$. Upon employing the representation [4]
\begin{equation}
\epsilon^n\cdot \dot{x}(t_n)]=\int d\xi_nd\bar{\xi}_n\,exp[i\xi_n\bar{\xi}_n\epsilon^n
\cdot \dot{x}(t_n)],
\end{equation}
where the $\xi_n,\bar{\xi}_n$ are Grassmann variables, we determine the Mth order term
in Eq. (18) to take the form
\begin{equation}
\left[\prod\limits_{n=M}^1ig(t^{c_n}_G)\right]^{ab}
\left[\prod\limits_{n=M}^1\int d\xi_nd\bar{\xi}_n\int\limits_s^{s'}dt_n\theta(t_n-t_{n-1})
\right]exp\left[i\sum\limits_{n=1}^M \hat{k}(t_n)\cdot x(t_n)\right]+permutations.
\end{equation}
In the above expression we have set
\begin{equation}
\hat{k}_\mu(t_n)\equiv p_{n,\mu}+\xi_n\bar{\xi}_n\epsilon^n_\mu \frac{\partial}{\partial t_n},
\end{equation}
while the term $`permutations$' refers to all possible rearrangements of the $t_n$.

As shown in Refs [11,12], via the employment of the Migdal-Polyakov area derivative [14,15], 
for a number of M background gluon field punctures of a given loop configuration one 
obtains the following result for the spin factor
\begin{eqnarray}
&&\Phi_{\mu\nu}^{[1]}[C]\rightarrow\Phi _{\mu \nu }^{[1]}[M]= 
{\rm P}\exp \left[ {\frac{i}{2}
\sum\limits_{n = 1}^M {J \cdot \phi (n)} } \right]_{\mu \nu } 
= \delta _{\mu \nu }  + \frac{i}{2}(J_{\rho \sigma } )_{\mu \nu } 
\sum\limits_{n = 1}^M{\phi_{\rho\sigma}}(n)\nonumber\\&& \quad +\left(\frac{i}{2}\right)^2 
(J_{\rho _2 \sigma _2 })_{\mu \lambda } (J_{\rho _1 \sigma _1 } )_{\lambda \nu } 
\sum\limits_{n_2  = 1}^M {\sum\limits_{n_1  = 1}^{n_2- 1}{\phi _{\rho _2 \sigma _2 }(n_2 )
\phi _{\rho _1 \sigma _1 } (n_1 ) + ...} }   
 \end{eqnarray}
with                   
\begin{equation}
\phi _{\mu \nu } (n) = 2\xi _n \bar \xi _n  (\varepsilon _\mu ^n p_{n,\nu }  - 
\varepsilon _\nu ^n p_{n,\mu } ) + \frac{4}{T}\xi _{n + 1} \bar \xi _{n + 1}  
\xi _n \bar \xi _n (\varepsilon _\mu ^{n + 1} \varepsilon _\nu ^n  - 
\varepsilon _\nu ^{n + 1} \varepsilon _\mu ^n )\delta (t_{n + 1}  - t_n )
\end{equation}
and where we have set $\bar{\xi}_{M+1}=\xi_{M+1}=0$.

For the generic partitioning, wherein we keep the first $N$ terms 
($0\leq N\leq M$) from the Wilson
loop expansion of the contour $C^{s_10}$ and the first $M-N$ terms from the contour 
$C^{ss_1}$, we obtain the following result for the two loop configuration with the four-vertex 
\begin{eqnarray}
&&\Gamma_{4,(M+2)}^{(2)}={1\over 8}g^M\sum\limits_{N=0}^M\left[\prod
\limits_{n=M}^{N+1}(t^{a_n}_G)\right]^{bc}(V_4)^{bcde}_{\mu\nu\rho\sigma}
\left[\prod\limits_{n=N}^1(t^{a_n}_G)\right]^{de}\left[\prod\limits_{n=M}^1
\int d\xi_n d\bar{\xi}_n\right]\nonumber\\&&\times \int\limits_0^\infty ds
\int\limits_0^s ds_1 \left[\prod\limits_{n=M}^{N+1}\int\limits_{s_1}^s\,dt_n\right]
\left[\prod\limits_{n=N}^1\int\limits_0^{s_1}dt_n\right]
\left[\prod\limits_{n=M}^1\theta (t_n-t_{n-1})\right]
\nonumber\\&&\quad\quad\times\Phi _{\mu \nu }^{[1]}[M-N]
\Phi _{\rho\sigma}^{[1]}[N]Q^{(1)}[\{\epsilon\},\{p\}]+permutations,
\end{eqnarray}
where we have denoted (the indication `(1)' signifies the presence of a single vertex)
\begin{equation}
Q^{(1)}=\int Dx(t)\delta[x(0)-x(s)]\delta[x(s_1)-x(s)]exp\left(-{1\over 4}
\int\limits_0^s\,dt_i\,\dot{x}^2+\sum\limits_{n=1}^M
\hat{k}(t_n)\cdot x(t_n)\right).
\end{equation} 

For the three-vertex, two-loop gluon configuration and keeping
$M-N_2$ terms from $C^{ss_2}$, $N_2-N_1$ 
from $C^{s_2s_1}$ and the remaining $N_2-N_1$ from $C^{s_10}$ we get
\begin{eqnarray}
&&\Gamma_{2,M+2)}^{3gluons}=-{1\over 8}g^M\sum\limits_{N_1=0}^M\sum\limits_{N_2=0}^M
\left[\left(\prod\limits_{n=M}^{N_2+1}t^{a_n}_G\right)t^b_G\left(
\prod\limits_{n=N_2}^{N_1+1}t^{a_n}_G\right)t^a_G
\left(\prod\limits_{n=N_1}^1t^{a_n}_G\right)\right]^{ab}\nonumber\\&&\times
\left[\prod\limits_{n=M}^1
\int d\bar{\xi}_n d\xi_n\right]\int\limits_0^\infty ds
\int\limits_0^s ds_2\int\limits_0^{s_2}ds_1 \left[\prod\limits_{n=M}^{N_2+1}
\int\limits_{s_2}^s\,dt_n\right]\left[\prod\limits_{n=N_2}^{N_1+1}
\int\limits_{s_1}^{s_2} dt_n\right]\left[\prod\limits_{n=N_1}^1
\int\limits_0^{s_1}dt_n\right]\nonumber\\&&\times
\left[\prod\limits_{n=M}^1\theta (t_n-t_{n-1})\right]V_3\left[N_1,N_2,-i\frac{\delta}
{\delta j}\right]R^{(2)}[\{\epsilon\},\{p\},j]_{j=0}+permutations,
\end{eqnarray}
where the current source $j$ has been introduced for the purpose of facilitating 
the computation of correlators of the type $<\dot{x}_\mu(t)\dot{x}_\nu(t')>$ 
which invariably enter the three-vertex computation. Finally,we have set
\begin{equation}
R^{(2)}(j)=\int Dx(t)\delta[x(s)-x(0)]\delta[x(s_1)-x(s_2)]exp\left(-{1\over 4}
\int\limits_0^s\,dt\,\dot{x}^2+i\sum\limits_{n=1}^M
\hat{k}(t_n)\cdot x(t_n)+i\int\limits_0^s dt\,j\cdot \dot{x}(t_n)\right )
\end{equation}
with the `(2)' serving to note that the configuration being considered has two vertex points.
The ghost-containing configuration results from Eq (28) via the 
replacement $V_3\rightarrow V_g$.

We are now at a point where we must execute the path integrals. Their gaussian nature
calls for determining the appropriate Green's functions on each given line contour.
It is possible to generalize the 
Green's functions corresponding to propagation of bosonic `particle' modes on 
unobstructed closed, or open, contours [4-6,12] in (Euclidean) space-time 
to our two-loop situation where one or two intervening vertex points are present. For loops 
with $n$ vertices the Green's function we need can be defined as follows
\begin{equation}
G^{(n)}(t,t')= G^{(n-1)}(t,t')-{1\over G^{(n-1)}(s_n,0)} [G^{(n-1)}(t,0)-G^{(n-1)}(t,s_n) 
-G^{(n-1)}(t',0)+ G^{(n)}(t',s_n)]^2, 
\end{equation}
with $G^{(0)}(t,t')={1\over s}|t-t'|(s-|t-t'|)$.

The corresponding expression for contours with non-coinciding initial and final points
(open contours), needed for the three-vertex configuration, reads
\begin{equation}
\tilde{G}^{(n)}(t,t')= G^{(n)}(t,t')+{1\over s_R^{(n)}}(t_R^{(n)}-t_R^{'(n)})^2
\end{equation}
with $\tilde{G}^{(o)}(t,t')=|t-t'|$.

In the above equation $s_R^{(n)}$ is the reduced total time given by
\begin{equation}
{1\over s_R^{(n)}}= {1\over s_1}+{1\over s_2-s_1}  +\cdot\cdot\cdot {1\over s -s_n}
\end{equation}
and $t_R^{(n)}$ is the reduced partial time
\begin{equation}
t_R^{(n)}=s_R^{(n)}\left\{\theta(s_1-t) {t\over s_1}  +[\theta(s_2-t)-\theta(s_1-t)]
{s_2-t\over s_2-s_1}  +\cdot\cdot\cdot+\theta(t_n-s){t-s_n\over s-s_n}\right\}.
\end{equation}  

Using the above expressions one arrives at the following results:
\begin{equation}
Q^{(1)}=(2\pi)^4\delta\left(\sum\limits_{n=1}^Mp_n\right)\frac{1}{(4\pi)^D[s(s-s_1)]^{D/2}}
exp\left[\sum\limits_{m>n}\hat{k}(t_n)\cdot\hat{k}(t_m)
G^{(1)}(t_n,t_m)\right]
\end{equation}

and
\begin{eqnarray}
R^{(2)}(0)&=&(2\pi)^4\delta\left(\sum\limits_{n=1}^Mp_n\right)\frac{1}{(4\pi)^D
[s_2(s-s_2)+s_1(s_2-s_1)]^{D/2}}\nonumber\\&&\times
exp\left[\sum\limits_{m>n}\hat{k}(t_n)\cdot\hat{k}(t_m)\tilde{G}^{(2)}(t_n,t_m)\right].
\end{eqnarray}

For the three vertex computation we need velocity correlators. The result obtains from
the operation involving the current source $j$ entering Eq (28). One obtains, generically,
\begin{eqnarray}
<\dot{x}_\mu(t)\dot{x}_\nu(t')>&=&-\delta_{\mu\nu}\partial_t\partial_{t'}
\tilde{G}^{(2)}(t,t')-\sum\limits_{n,m}\hat{k}(t_n)\hat{k}(t_m)  
\partial_t[\tilde{G}^{(2)}(t,0)-\tilde{G}^{(2)}(t,t_n)]
\nonumber\\&&\quad\quad\times\partial_{t'}[\tilde{G}^{(2)}(t',0)-\tilde{G}^{(2)}(t',t_m)].
\end{eqnarray}

Substituting $Q^{(1)}$, $R^{(2)}$ and the corresponding correlators in Eqs (26) 
and (28), respectively
(the latter in the `ghost' term as well) a set of master expressions
is produced which call for
the execution of integrations over two sets of parameters. We have been able to reproduce
known results [16] for the $M=0$ case and $M=2$ cases (forthcoming paper).   

For our closing comments let us begin by saying that the ability to reformulate QCD in terms
of space-time paths has led to the result of producing master expressions, given in terms
of two sets of parameters (Grassmann and Feynman) which correspond to multi-gluon loop 
configurations that enter perturbative expansions. Once the overall strategy is established
the only real determination that needs to be made is the computation of Green's functions
describing bosonic propagation on paths with a number of `marked' points satisfying two 
types of boundary conditions. Even though the examples we have discussed pertain to paths
that can be continuously traced nothing, in principle, prevents one from treating
more complex, higher loop, configurations which call for more than a single continuous
parametrization. The bottom line is that the relevant path integral, whose execution
calls for determining the Green's functions, is Gaussian, therefore manageable.

Aside from the execution of multi-gluon 
loop computations, which call for numerical approaches to confront the relevant 
master expressions as the configurations get increasingly complicated, there is another
aspect of the Polyakov path integral approach to QCD that we find more intriguing. This
has to do with the fact that each worldline, in addition to being geometrically 
characterized by the spin of the propagating particle mode through the
spin factor, is accompanied by a `cloud' of background gauge potentials on account of 
the Wilson line(loop). Consider a context within which the background fields are quantized
in a way that they carry non-perturbative physics. This, e.g., can be achieved in a lattice
context or, alternatively, in lattice inspired continuum schemes, as in Refs [17-19]. One
would then be in possession of a methodology formulated strictly within the bounds
of QCD that will facilitate the study of the interplay between the perturbative
and the non-perturbative regime of the theory. It is to the confrontation of this problem
that we intend to direct our future efforts.

\end{document}